\documentclass[11pt,aps,preprint,floats,nofootinbib]{revtex4}

\usepackage{color} 
\usepackage{graphicx}
\usepackage{ifthen}
\usepackage{epsf}

\def\mem#1{\mbox {\underline{#1}}} 

\def\dert#1{{ d #1 \over d t}}

\newcommand{\nc}{\newcommand}
\nc{\postscript}[2] 
{\setlength{\epsfxsize}{#2\hsize}\centerline{\epsfbox{#1}}}
\nc{\bg}{B. Grzadkowski}
\nc{\non}{\nonumber}
\nc{\barx}{\bar{x}}\nc{\pbarn}{\;\hbox {pb}}\nc{\fbarn}{\;\hbox {fb}}
\nc{\vtrue}{v_0}
\nc{\vtree}{v}
\nc{\veff}{V_{\rm eff}}
\nc{\hc}{\hbox {h.c.}} 
\nc{\re}{\hbox {Re}} 
\nc{\im}{\hbox {Im}}
\nc{\mev}{\hbox {MeV}} 
\nc{\gev}{\;\hbox {GeV}} 
\nc{\tev}{\;\hbox {TeV}}
\def\gesim{\lower0.5ex\hbox{$\:\buildrel >\over\sim\:$}} 
\def\lesim{\lower0.5ex\hbox{$\:\buildrel <\over\sim\:$}} 
\nc{\xprd}[3]{{\it Phys.\ Rev.}\ {{\bf D{#1}} (#2), #3}}
\nc{\xprb}[3]{{\it Phys.\ Rev.}\ {{\bf B{#1}} (#2), #3}}
\nc{\xprl}[3]{{\it Phys.\ Rev.\ Lett.}\ {{\bf {#1}} (#2), #3}}
\nc{\pr}[3]{{\it Phys.\ Rep.}\ {{\bf {#1}} (#2), #3}}
\nc{\plb}[3]{{\it Phys.\ Lett.}\ {{\bf B{#1}} (#2), #3}}
\nc{\npb}[3]{{\it Nucl.\ Phys.}\ {{\bf B{#1}} (#2), #3}}
\nc{\ptp}[3]{{\it Prog.\ Theor.\ Phys.}\ {{\bf {#1}} (#2), #3}}
\nc{\zfp}[3]{{\it Z.\ Phys.}\ {{\bf C{#1}} (#2), #3}}
\nc{\mpla}[3]{{\it Mod.\ Phys.\ Lett.}\ {{\bf A{#1}} (#2), #3}}
\nc{\xrmp}[3]{{\it Rev.\ Mod.\ Phys.}\ {{\bf {#1}} (#2), #3}}
\nc{\ijmpa}[3]{{\it Int.\ J.\ Mod.\ Phys.}\ {{\bf A{#1}} (#2), #3}}
\nc{\jhep}[3]{{\it JHEP}\ {{\bf #1} (#2), #3}}
\nc{\etal}{{\it et al.}}
\def\lsim{\mathrel{\raise.3ex\hbox{$<$\kern-.75em\lower1ex\hbox{$\sim$}}}}
\def\gsim{\mathrel{\raise.3ex\hbox{$>$\kern-.75em\lower1ex\hbox{$\sim$}}}}

\nc{\lspace}{\;\;\;\;\;\;\;\;\;\;} \nc{\llspace}{\lspace \lspace}

\nc{\beq}{\begin{equation}}  \nc{\eeq}{\end{equation}}
\nc{\bea}{\begin{eqnarray}}  \nc{\eea}{\end{eqnarray}}
\nc{\baa}{\begin{array}}     \nc{\eaa}{\end{array}}
\nc{\bit}{\begin{itemize}}   \nc{\eit}{\end{itemize}}
\nc{\ben}{\begin{enumerate}} \nc{\een}{\end{enumerate}}
\nc{\bce}{\begin{center}}    \nc{\ece}{\end{center}}

\nc{\mh}{m_h}
\nc{\mt}{m_t}
\nc{\mz}{m_Z}
\nc{\mw}{m_W}
\nc{\la}{\lambda}
\nc{\La}{\Lambda}

\def\half{\frac12}
\def\lcal{{\cal L}}
\def\up#1{^{(#1)}}
\def\inv#1{\frac1{#1}}
\def\ocal{{\cal O}}
\def\pb{\bar\varphi}

\nc{\al}[1]{
\ifthenelse{\equal{#1}{p}}{\alpha_{\phi}}{}
\ifthenelse{\equal{#1}{dp}}{\alpha_{\partial\phi}}{}
\ifthenelse{\equal{#1}{p1}}{\alpha_{\phi}^{(1)}}{}
\ifthenelse{\equal{#1}{p3}}{\alpha_{\phi}^{(3)}}{}
\ifthenelse{\equal{#1}{tp}}{\alpha_{t\phi}}{}
\ifthenelse{\equal{#1}{qt1}}{\alpha_{qt}^{(1)}}{}
\ifthenelse{\equal{#1}{ll3}}{\alpha_{ll}^{(3)}}{}
\ifthenelse{\equal{#1}{pl3}}{\alpha_{\phi l}^{(3)}}{}
}

\begin{document}

\baselineskip 15pt

\def\mib#1{\mbox{\boldmath $#1$}}
\def\bra#1{\langle #1 |} \def\ket#1{|#1\rangle}
\def\vev#1{\langle #1\rangle} \def\dps{\displaystyle}

\begin{flushright}
IFT-18/2003\\
UCRHEP-T359\\
July, 2003
\end{flushright}

\title{\Large\bf Higgs-Boson Mass Limits and Precise Measurements\\
beyond the Standard Model}

\author{\sc Bohdan Grzadkowski,$^{\:a)\:}$ Jacek
Pliszka,$^{\:a),\:b)\:}$ Jos\'e Wudka$^{\:b)\:}$} 
\vspace{1cm}

\affiliation{{\sl a)} Institute of Theoretical Physics, Warsaw University,
 Ho\.za 69, PL-00-681 Warsaw, POLAND}

\affiliation{{\sl b)} Department of Physics, University of California, 
Riverside CA 92521-0413, USA}

\date{\today}

\begin{abstract}
The triviality and vacuum stability bounds on the Higgs-boson mass ($\mh$)
were revisited
in presence of  weakly-coupled new interactions  parameterized
in a model-independent way by effective operators of dimension 6.
The constraints from precision tests of the Standard Model were taken
into account.
It was shown that for the scale of new physics in the region 
$\La \simeq 2 \div 50 \tev$
the Standard Model triviality upper bound remains unmodified
whereas it is natural to expect that 
the lower bound derived from the requirement of vacuum stability is
substantially modified depending on the scale $\La$ and strength
of coefficients of effective operators. 
A natural generalization of the standard triviality
condition leads also to a substantial reduction of the allowed region in 
the ($\La,\mh$) space.
\end{abstract}


\maketitle


\section{Introduction}
\label{sect:intro}

In spite of a huge experimental effort, the Higgs particle, the last 
missing ingredient of 
the Standard Model (SM) of electroweak interactions has not been yet discovered. 
For a Higgs-boson mass $\lesim 115 \gev$ the most promising production channel
at LEP2 is through radiation off a $Z$-boson: $ e^+ e^- \to Z h $; using this 
reaction the  LEP experiments obtained the limit~\cite{higgs_limit} 
$ \mh > 113.2 \gev $ on the SM Higgs-boson mass.
The Higgs particle also contributes radiatively to several well measured
quantities, which has been used to derive the complementary
upper bound~\cite{prec_data} $\mh \lesim 212 \gev$ at 95 \%~C.L.. 
The data then leave a rather narrow range for $\mh$, however it must be emphasized 
that these
constraints are highly model-dependent and are significantly
weakened in most extensions of the SM .

There also exist theoretical, restrictions of $\mh$ based on the so-called
triviality and vacuum stability arguments. As it is well know~\cite{triviality} 
the renormalized $\phi^4$ theory cannot contain an interaction term ($\la \phi^4$) for
any non-zero scalar mass: the theory must be trivial. Within a 
perturbative approach the statement corresponds to the fact
that for any non-zero scalar mass~\footnote{Since
the (tree-level) mass is $ \propto \sqrt{\la}$ this condition corresponds to a
non-vanishing  initial value for the renormalization group (RG)
evolution of $\la$.} there exists a finite energy scale at 
which $\la$ diverges (the Landau pole). Consequently this
theory is consistent for all energy scales only when it describes
non-interacting scalars. An analogous effect 
occurs in the scalar sector of the SM, though modified to some extent by presence
of gauge and Yukawa interactions. This, however,
does not necessarily imply a trivial scalar sector, since we do not demand
the validity of the SM at arbitrarily high energy scales. For
example, it is often assumed that the SM represents the
low energy limit of some underlying more fundamental theory whose heavy
excitations decouple at low energy, but become manifest at a scale $\La$.
Within this scenario the SM is an effective theory containing the dominant 
terms in a $1/\La$ expansion; any process occurring at a typical
energy $E$ will then receive corrections suppressed by powers of $E/\La$
generated by the sub-leading interactions.

If the SM is to be accurate for energies below $ \La $ the Landau pole
should occur at scale $ \La $ or above, and this condition gives a
($\La$-dependent) upper bound on $ \mh $~\cite{triv_bounds}. 
On the other hand, for sufficiently small $ \mh $ radiative corrections can
destabilize the ground state. This occurs if the running scalar
self-coupling constant $ \la $ becomes negative  at some scale, that
can be again identified with the scale of new physics $\La$. Alternatively
requiring the SM vacuum to be stable for scales below $ \La$ implies a
($\La$-dependent) lower bound on $ \mh $~\cite{vacuum_bounds}. 

The consequences of the above arguments (triviality and vacuum stability) are 
usually discussed assuming pure SM interactions. However,
if the scale of new physics is sufficiently low (of the order of a few TeV)
one would expect for the sub-dominant effects to significantly influence
both the renormalization group evolution and the scalar effective potential,
and thus modify the corresponding bounds on the Higgs-boson mass.

It then becomes interesting to determine the manner in which heavy
physics with scales in the $ 10 \tev $ region can modify the stability and
triviality bounds on the Higgs-boson mass. In this paper we address this 
question in a model-independent way by parameterizing the heavy physics
effects using  an effective Lagrangian satisfying the SM gauge
symmetries. This issue was investigated in previous 
publications~\cite{Grzadkowski:2001vb}, but   without
taking into account the restrictions generated by the precision tests of the SM.
The analysis presented here remedies this deficiency
by including these constraints (to the one-loop approximation
in the SM and at tree level in the effective dim~6 operators); for other works discussing 
vacuum stability including some effective operators see 
Refs.~\cite{Datta:1996ni,Casas:2000mn,Burgess:2002tj}.

The paper is organized as follows. In Sec.~\ref{non_stand}, we define the Lagrangian
relevant for our discussion. Sec.~\ref{rge} presents the relevant renormalization group
running equations including effects of non-standard interactions.
In Sec.~\ref{effpot.sec} we calculate the effective potential with one insertion 
of an effective operator. Sec.~\ref{strategy} contains the methodology that we have
applied and our numerical results.
Concluding remarks are given in Sec.~\ref{summary}.

\section{Non-Standard Interactions}
\label{non_stand}

Our study of the stability and triviality constraints on the Higgs-boson mass
will be based on the SM Lagrangian modified by the addition of a series
of effective operators whose coefficients parameterize the low-energy 
effects of the heavy physics~\cite{leff.refs}. 
Assuming that these non-standard effects
decouple implies~\cite{decoupling} that all physical effects disappear in the 
$ \La \to \infty$ limit and, in particular, that
 the effective operators of dimension $>4$ appear multiplied
 by appropriate inverse powers of $ \La $.
Leading effects are then
generated by operators of mass-dimension 6 (dimension 5 operators
necessarily violate lepton number~\cite{effe_oper} and are presumably
associated with new physics at
very large scales since they lead to very small effects; accordingly they can 
be safely ignored hereafter).
Given our emphasis on Higgs-boson physics the effects of all fermions
excepting the top-quark can also be 
ignored~\footnote{We assume that the masses are natural in the technical sense~\cite{thooft},
so that effective couplings containing the Higgs boson and the light fermions are 
suppressed by powers of the corresponding Yukawa couplings.}.
We then have
\bea
\lcal_{\rm tree}& = & 
-\frac14 F_{\mu\nu}^iF^{i\mu\nu} -\frac14 B_{\mu\nu}B^{\mu\nu}+ 
\left| D \phi \right|^2  - 
\la \left(|\phi|^2 -\half v^2 \right)^2 + \non \\ 
&& i \bar q \not\!\!D q +
i \bar t \not\!\!D t + 
f \left( \bar q \tilde\phi t + \hbox{h.c.} \right) +
\sum_i\frac{\alpha_i}{\La^2}{\cal O}_i,
\label{lagrangian}
\eea
where $\phi$ ($\tilde \phi = -i \tau_2 \phi^* $), 
$q$ and $t$ denote the scalar doublet, third generation 
left-handed quark doublet and the right-handed top singlet, respectively. 
$D$ represents a covariant derivative, and $F_{\mu\nu}^i$ and $B_{\mu\nu}$ the 
$SU(2)$, $U(1)$ field strengths whose corresponding gauge couplings we denote 
by $g$ and $g'$. The factors $\alpha_i$
are unknown coefficients that parameterize the low-energy effects of
the non-standard interactions and we have neglected contributions $
\propto 1/\La^4 $. 

For weakly coupled theories, the 
$ \alpha_i $ that can be generated only through loop effects are
sub-dominant as they are suppressed by numerical factors $ \sim 1/
(4\pi)^2 $~\cite{tree_oper}; hence we will consider only those
operators that can be generated at tree-level by the heavy physics. Even
with all the above restrictions there remain 16 operators that 
involve exclusively the fields in (\ref{lagrangian}). Of these only
5 contribute directly to the effective potential, the remaining 11 would
affect our results only through their RG mixing which, being suppressed by
a factor $ \sim G_F \La^2 $ (where $G_F$ denotes the Fermi constant)
are expected to play a sub-dominant role.
In the calculations below we will include only one of these operators
for illustration purposes;
our results justify the claim that the corresponding effects are
small.

Specifically we include the following set of operators:
\beq
\baa{lll}
{\ocal_{\phi}} = \inv3 | \phi|^6 &
{\ocal_{\partial\phi}} = \half \left( \partial | \phi |^2 \right)^2 &
{\ocal_{\phi}\up1} = | \phi |^2 \left| D \phi \right|^2 \cr
{\ocal_{\phi}\up3} = \left| \phi^\dagger D \phi \right|^2 &
{\ocal_{t\phi}} = | \phi |^2 \left( \bar q \tilde\phi t + \hbox{h.c.} \right) &
{\ocal_{qt}\up1} = \half \left|\bar q t \right|^2
\eaa
\label{operators}
\eeq
where
${\ocal_{\phi}}$, ${\ocal_{\partial\phi}}$, ${\ocal_{\phi}\up1}$, ${\ocal_{\phi}\up3}$,
${\ocal_{t\phi}}$ are the 5 operators contributing to the effective
potential, while ${\ocal_{qt}\up1}$ is included to estimate the effects
of RG mixing.

Of the first five operators only  ${\ocal_{\phi}} = \inv3 | \phi|^6$ contributes
at the tree level to the scalar potential:
\beq
V\up{\rm tree}= 
- \eta \La^2 |\phi|^2  + \la |\phi|^4  -
{\alpha_{\phi} \over 3 \La^2 } | \phi|^6
\label{tree_pot}
\eeq
where we have used the notation: $\eta \equiv \la v^2/\La^2 $.

\section{The renormalization group equations}
\label{rge}

In order to test the high energy behavior of the scalar potential one has to derive
the RG running equations for $\la$, $\eta$ and $\alpha_{\phi}$.
The $\beta$ functions for these parameters are
influenced by all the operators in (\ref{operators})  
and by the gauge and Yukawa interactions, so the full RG evolution also
require the $ \beta $ function for the corresponding couplings.
In the following calculations we will
adopt dimensional regularization and $\overline{\rm MS}$ 
renormalization scheme. We will restrict ourselves to the
one-loop approximation keeping 
all SM contributions as well as those linear in 
the effective operators (\ref{operators}). 

Defining $ \bar\alpha = \alpha_{\partial\phi} +2
\alpha_\phi\up1 + \alpha_\phi \up3 $,   the resulting 
evolution equations are: 
\def\dert#1{{ d #1 \over d t}}
\begin{eqnarray}
\dert \lambda &=&12\lambda^2 -3 f^4 + 6 \lambda f^2 
-{3\over2}\la \left(3 g^2 + g'{}^2 \right)
+{3\over16} \left(g'{}^4 +2 g^2 g'{}^2 + 3 g^4\right)
\cr &&
+ 2 \eta \left[2 \alpha_\phi +
\lambda \left( 3 \alpha_{\partial\phi} +4 \bar\alpha + \alpha_\phi\up3\right) \right]
\cr
\dert \eta &=& 3\eta\left[2\lambda + f^2 - {1\over4}\left( 3g^2+ g'{}^2 \right)\right]
+2 \eta^2  \bar\alpha
\cr
\dert f &=& {9\over4} f^3 -f \left(4 g_s^2 +  {9\over8} g^2 + {17\over24} g'{}^2 \right) 
- 3 \eta\alpha_{t\phi} + {f\eta\over2} \left(\bar\alpha + 3 \alpha_{qt}\up1 \right)  \cr
\dert{\alpha_\phi}&=&
 9 \alpha_\phi \left(6 \lambda + f^2 \right)
 + 12 \lambda^2 (9\alpha_{\partial\phi}+6 \alpha_\phi\up 1
 +5\alpha_\phi\up 3)
 + 36 \alpha_{t\phi} f^3 - {9\over4} \left( 3 g^2 + g'{}^2 \right) \alpha_\phi
\cr &&
- {9\over8}\left[2  \alpha_\phi\up1 g^4 + \left(\alpha_\phi\up1 +
\alpha_\phi\up3 \right)\left(g^2 + g'{}^2 \right)^2 \right] \cr
\dert{\alpha_{\partial\phi}}&=& 2 \lambda
 \left( 7 \alpha_{\partial\phi} - \alpha_\phi\up1 + \alpha_\phi \up3
 \right)
 + 6 f \left(f \alpha_{\partial\phi}  - \alpha_{t\phi}\right) \cr
\dert{\alpha_\phi\up1}&=& 2 \lambda
 \left(\bar\alpha+3\alpha_\phi\up1\right)
 + 6 f \left(f \alpha_\phi\up1
 -\alpha_{t\phi}\right) \cr
\dert{\alpha_\phi\up3}&=& 6 (\lambda +f^2) \alpha_\phi\up3 \cr
\dert{\alpha_{t\phi}} &=& -3 f (f^2+\lambda) \alpha_{qt}\up1
+ \left( {15\over4}f^2-12\lambda\right) \alpha_{t\phi}
- {1\over2}f^3 \left(
2\alpha_{\partial\phi} +\alpha_\phi\up1 + \alpha_\phi \up3 \right) \cr
\dert{\alpha_{qt}\up1} &=& {3\over2} \alpha_{qt}\up1 f^2
\label{beta_fun}
\end{eqnarray}
where $8 \pi^2 t\equiv \log (\kappa/\mz)$,
$ \kappa $ denotes the renormalization scale and $g_s$ is the QCD coupling constant. 
The RG equations for 
$g$, $g'$, and $g_s$; $dg/dt = -19 g^3/12$, $dg'/dt = + 41 g'{}^3/12 $, and $ d g_s/dt = -7/2 g_s^3$
are not modified by the $ \alpha_i$.

From this set of equations it is straightforward to obtain the traditional
triviality constraints on
$\mh$ as a function of $\La$ by requiring that the position of the Landau pole
in the evolution of $\la(t)$ lies beyond the scale $\La$. 
At this point it is important to note that within perturbation theory the
triviality constraint is {\em not} obtained from the requirement that the Higgs mass
diverges at scale $ \La$, but form the condition that the theory remains perturbative 
at all scales below $ \La $. The triviality
bound on $\mh$ will be obtained then by requiring $\la$ {\it and} $|\alpha_i|$ to remain
below certain specified values (chosen so as to insure perturbative
consistency) up to the scale $\La$; the details are presented 
in Sect.\ref{strategy}

In order to solve the equations (\ref{beta_fun}) we have to specify
appropriate boundary conditions.
For the SM parameters these are determined by requiring that the correct
physical parameters (such as the Higgs-boson mass, top-quark masses, etc.)
are obtained at the electroweak scale, and
that the correct SM ground state is realized; the details 
of the implementation of these low scale 
initial conditions are also described in  Sect.\ref{strategy}. 
In contrast, the boundary conditions for the
$\alpha_i$ are naturally specified at the scale $\kappa=\La$
since it is below this scale that (\ref{lagrangian}) is expected
 to describe the effects of the heavy excitations; following Ref.~\cite{tree_oper} we will
use the (natural) choices $\alpha_i|_{\kappa=\La} = \pm 1$.

The triviality bound is obtained  by solving the equations (\ref{beta_fun}) with the mixed 
(defined partly at the electroweak scale $\mz$ and partly at
the new-physics scale $\La$) 
boundary conditions  described above and requiring that triviality constraints 
(see Sec.\ref{strategy}) are saturated, and this provides a relationship 
between $ \mh $ and $ \La $. For example, if we require only $ \lambda(\kappa) < \pi/2 $
for $ \mz \le \kappa \le \La $, we obtain the plot in 
Fig.\ref{triv-only-plot}; imposing also the additional conditions 
$ | \alpha_i ( \kappa ) | <3/2 $ yields a much richer structure discussed in Sec.\ref{strategy} and
illustrated by the plots in Fig.\ref{models-plot}.

\section{The effective potential}
\label{effpot.sec}

In order to investigate the vacuum structure of the effective theory we 
first calculate the effective potential:
\begin{equation}
V_{\rm eff} = - \sum_N \inv{N!} \Gamma\up N(0) \pb^n,
\label{eff_pot_def}
\end{equation}
where $\Gamma\up N(0)$ are N-point one-particle-irreducible Green`s functions
with zero external momenta and $\pb$ is the classical scalar field.
Adopting the Landau gauge~\footnote{As it has been noticed in Ref.~\cite{gauge_dep_eff_pot}
the effective potential (as a sum of off-shell Greens functions) is gauge dependent.
Therefore the bounds on the Higgs-boson mass 
derived from vacuum stability arguments
can depend on the gauge parameter adopted in the loop calculation~\cite{gauge_dep_bound}. 
However, since the $\beta$
functions and the tree-level potential $V_{\rm eff}\up{\rm tree}$ are 
gauge-independent,
a consistent RG improved tree-level effective potential is in fact gauge independent.
For the one-loop SM RG improved effective potential, the error caused by the gauge 
dependence
has been estimated in Ref.~\cite{quiros} at $\Delta \mh \lesim 0.5 \gev$.} 
we find:
\begin{eqnarray}
V_{\rm eff}(\pb) &=&
-\eta \La^2 |\pb|^2 + \la |\pb|^4 - {\alpha_\phi |\pb|^6\over3 \La^2} \\
&&+ {1 \over 64 \pi^2} \Biggl[ 
   H^2 \left( \ln {H \over \kappa^2} - {3\over2} \right) +
3  G^2 \left( \ln {G \over \kappa^2} - {3\over2} \right) +
6  W^2 \left( \ln {W \over \kappa^2} - {5\over6} \right) \cr
&&+3  Z^2 \left( \ln {Z \over \kappa^2} - {5\over6} \right) -
12 T^2 \left( \ln {T \over \kappa^2} - {3\over2} \right)
- 4\eta^2\La^4 \left( \ln {\eta \La^2 \over\kappa^2} - {3\over2} \right)\Biggr], \non
\label{effpot}
\end{eqnarray}
for 
\begin{eqnarray}
H &=& (6 \la |\pb|^2 - \eta \La^2) - \left[(6 \la |\pb|^2 - \eta \La^2)
(2 \alpha_{\partial\phi} + \alpha_\phi\up1 + \alpha_\phi\up3) + 
5 \alpha_\phi |\pb|^2 \right] {|\pb|^2\over \La^2} \cr
G &=& (2 \la |\pb|^2 - \eta \La^2) -  \left[(2 \la |\pb|^2 - \eta \La^2)
\left(\alpha_\phi\up1 + \inv3\alpha_\phi\up3 \right)  + \alpha_\phi |\pb|^2
\right] {|\pb|^2\over \La^2} \cr
W &=& {g^2  \over2}|\pb|^2 \left( 1 + { |\pb|^2 \alpha_\phi\up1 \over \La^2 } \right) \cr
Z &=& { g^2 +g'{}^2 \over2} |\pb|^2  \left( 1 + { |\pb|^2 ( \alpha_\phi\up1 + 
\alpha_\phi\up3)   \over \La^2 } \right) \cr
T &=& f^2  |\pb|^2 \left(1 + { 2 \alpha_{t \phi} | \pb|^2 \over f \La^2} \right),
\end{eqnarray}
where, as mentioned above, $g$ and $g'$ denote respectively the 
$SU(2)$ and $U(1)$ running gauge coupling constants.
The {\em form} of the effective potential is precisely the same as the
one in the pure SM, the whole effect of the effective operators can be
absorbed in a re-definition of the quantities $H$, $G$, etc.~\footnote{This
expression
(to the leading order in the $\alpha_i$)
for the effective potential was obtained following
the usual diagrammatic approach (with
one insertion of each effective operator) according
to (\ref{eff_pot_def}); identical results were derived
using the functional definition of the effective potential~\cite{func_def}.}  
It should be noticed here that the last ($\pb$-independent) term in (\ref{effpot})
is needed to insure that $ V_{\rm eff} $ is scale invariant (for details see \cite{cosm_const}).
This, however, does not determine this contribution uniquely; our choice also insures
 $V_{\rm eff}(\pb=0)=0$ as implied by the diagrammatic definition (\ref{eff_pot_def}).

Since we will consider values of $\pb$ substantially larger than the electroweak
scale $\vtrue $, we shall chose an appropriate renormalization scale 
$\kappa \sim \pb$ in order
to moderate the logarithms that appear in the effective potential.
As in the previous section we shall use the RG running equations
to relate the coupling constants
renormalized at the  scale $\kappa = \pb$
to the various input  parameters.

Finally (and unlike the pure $\phi^4$ theory), it is worth noting that 
the interaction of the scalars
with the fermions and gauge bosons, generate a non-trivial
scalar field anomalous dimension $\gamma$ at the one-loop level. 
We therefore also include the
corresponding scale dependence of $ \pb $ (for details see~\cite{Grzadkowski:2001vb}).
Hereafter we will consider the RG improved effective potential $ V_{\rm
eff} ( \pb(t)) $ that includes all these effects.

\section{Strategy and numerical results}
\label{strategy}

In considering stability and triviality limits we studied 
models characterized by having $ | \alpha_i(\La)| = 1 $
and $\la,~f,~\eta,~g$ and $g'$ at $ \kappa = \mz$
consistent with a choice of $ \mh$ and $ \La$
and with the experimental values of $ \mt,~ \mw,~ \mz,$ and $ \rho $ (the relative
strength of neutral and charged currents).
For this we used the expressions~\cite{effe_oper,stu_effective_1}
\bea
\mh^2 &=& 2 \la \vtrue ^2\left[ 1 -
{\vtrue ^2 \over 4 \La^2 } \left(4\alpha_{\partial\phi} + 2 \alpha_\phi\up1+ 2 \alpha_\phi\up3+
{~ \alpha_\phi\over\la} \right) \right]+ \delta\up1 \mh^2, \cr
\mt &=& {\vtrue  \over \sqrt{2}}\left(f -
\alpha_{t\phi}{\vtrue ^2\over \La^2}\right) + \delta\up1 \mt, \cr
\mw &=& \frac12 g \vtrue \left[ 1 + {\eta \al{p1} \over 4 \la } \right] + \delta\up1 \mw \label{ht_mass}\\
\mz &=& \frac12 \sqrt{g^2 + g'{}^2} \vtrue \left[ 1 + {\eta \left( \al{p1} 
+ \al{p3} \right)\over 4 \la } \right] + \delta\up1 \mz \cr
\rho &=& 1 - {g'{}^2 \eta \al{p3} \over 2 g^2 \la } + \delta\up1 \rho \non
\eea
where $\delta\up1 \mt , ~ \delta\up1 \mh^2, $ etc. denote the one-loop SM radiative 
corrections~\cite{Aoki:ed,Hollik:1988ii,quiros}. 

We started by choosing a set of signs for the $ \alpha_i(\La) $, 
taking $f,~\eta,~g$ and $g'$ at $ \kappa = \mz$ equal to their tree-level values,
and making a choice of $ \mh$ and $ \La$ in the region $ 2 \tev \le \La \le 50 \tev,
~ 65\gev \le \mh \le 1 \tev $. Having thus specified the initial conditions, we
numerically solved the RG evolutions equations and
checked the numerical stability. If satisfactory, these solutions were
used to evaluate $ \mt,~ \mw,~ \mz,$ and $ \rho $ in (\ref{ht_mass}).
Taking then the experimental values and uncertainties we constructed
the combined $ \chi^2$ function for these 4 observables~\footnote{Since the
experimental uncertainties for $\mw$ and $\mz$ are smaller than the
theoretical counterparts within the 1-loop approximation, we used
1\%\ (theoretical) error for the correpsonding contributions to $ \chi^2$.}.
If $ \chi^2 > 25 $ was obtained, then the initial 
values of $f,~\eta,~g$ and $g'$ were adjusted until the results
yielded $ \chi^2 < 25$. Once this was achieved
the solution was deemed consistent with the precision measurements
and was used to determine the triviality and stability conditions
given the choices of $ \alpha_i,~\mh$ and $ \La $.

We found that the above procedure for determining  $f,~\eta,~g$ and $g'$ at $ \kappa = \mz$
failed  only when $ \La \lesim 5 \tev$ and $ \la > \pi/2 $, so that these cases are 
already disallowed by the triviality constraints (see below).

The requirements we use to implement the  triviality constraint are the following
\renewcommand{\labelenumi}{T\arabic{enumi}:}
\renewcommand{\labelenumii}{T\arabic{enumi}\alph{enumii}:}
\begin{enumerate}
\item $\lambda < \frac{\pi}{2}$, 
\item $|\alpha_i| < 1.5 $ for all $i$, 
\item logical product of the following 3 conditions:\begin{enumerate}
    \item $|\eta \alpha_i | < \frac{\la}{4}$,
    \item $|\frac{\al{p}}{\la} | < \frac{3}{4} |\frac\La\kappa|^2$,
    \item $|\eta (4 \al{dp} + 2 \al{p1} + 2 \al{p3}+ \al{p}/\la)|<|\la|$.
    \end{enumerate}
\end{enumerate}
\renewcommand{\labelenumi}{\arabic{enumi}}
\renewcommand{\labelenumii}{\alph{enumii}}
for all scales $ \mz \le \kappa \le \kappa_{\rm max} = (3/4)\La $ (all quantities represent the running
expressions obtained by solving (\ref{beta_fun})). We do not allow $ \kappa $
 to reach $ \La $ since the Lagrangian (\ref{lagrangian}) is valid
only below this scale; the specific choice of $ \kappa_{\rm max} $
is arbitrary and the results are not sensitive to it.

T1 and T2 are standard triviality conditions insuring that the coupling constants
remain small enough for perturbation theory to remain valid.
The condition T3 contains three parts: T3a,T3b and T3c that ensure 
that corrections from 6-dim operators remain small.
T3a guarantees that the non-standard corrections to the SM $\beta$-functions 
are below 25\% level (and is satisfied if T2 is fulfiled).
T3b keeps the $\phi^6$ effective contribution to the tree level potential (\ref{tree_pot})
small ($<$25\%) in comparison to the SM $\phi^4$ quartic term. Lastly,
T3c, requires for the effective operator corrections 
to the Higgs boson to be below 25\% (see eq.~\ref{ht_mass}).
T3 then implement the condition that the leading $ 1/\La$ effects remain small
compared to the SM contributions.

 The vacuum stability requirement is implemented by the following 2 conditions:
\renewcommand{\labelenumi}{S\arabic{enumi}:}\begin{enumerate}
\item For $ \pb \le \frac34\La$, $V_{\rm eff} (\pb)$ has a unique minimum at 
$ \pb = \vtrue $ within 20\% of the SM tree-level value $ \vtree\simeq 246$GeV,
\item The potential at $\pb = \frac34\La$ lies above its value at the minimum.
\end{enumerate}\renewcommand{\labelenumi}{\arabic{enumi}}
S1 implements the condition that the underlying theory is weakly coupled while
S2 insures that the minimum at $ \pb = \vtrue$ is stable for all field
strengths below $ \frac34\La$.

\subsection{Lower bound on the Higgs boson mass}

\begin{figure}
\centering%
\includegraphics[width=12cm]{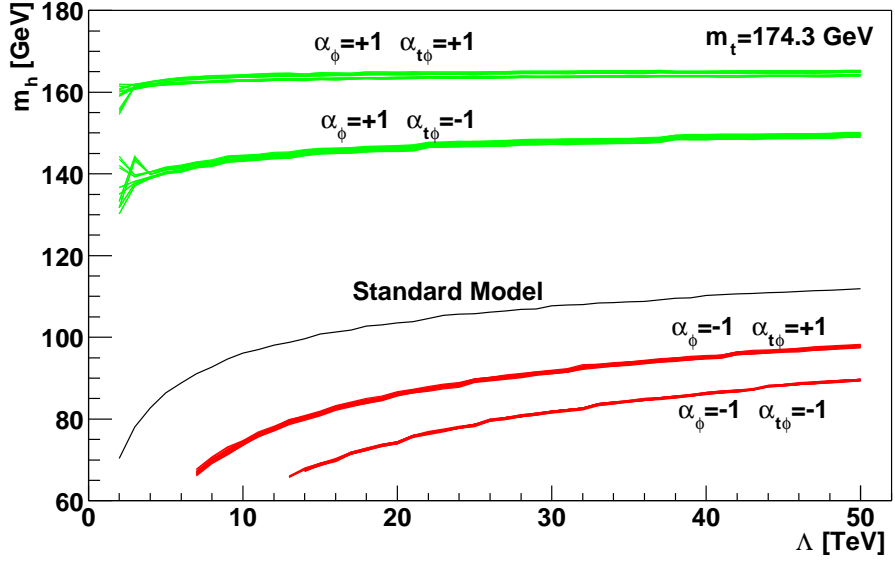}%
\caption{Lower bounds on the Higgs boson mass (all  S's conditions satisfied),
the black curve represents the SM limit, the upper (green) curves are for $\alpha_\phi>0$, 
and the lower (red) ones
for $\alpha_\phi<0$. For each color the 
higher branches correspond to $\alpha_{t\phi}<0$ while lower for $\alpha_{t\phi}>0$.}%
\label{stab-only-plot}%
\end{figure}

Due to its appearance in the tree-level potential ({\ref{tree_pot}), $\al{p}$ has a strong
impact on the vacuum stability bound. For 
$\al p(\La) > 0 $ the corresponding term decreases the value
of  $ V_{\rm eff} $ at $ \pb \sim \La $ and a {\em larger} value of $ \la $
is required to stabilize the minimum at $ \vtrue $
(thereby insuring condition S2 is satisfied).
In this case we then obtain that the vacuum stability constraints are satisfied
for values of $\mh$ larger than those obtained in the pure SM (for the same
choice of $ \La$).

When $ \al p (\La) < 0 $ the effect of the corresponding term in ({\ref{tree_pot})
tends to stabilize the minimum at $\vtrue$, but it also shifts it
away from the tree-level SM value. {\it Therefore in this case $\mh$ is not limited
by the stability of $\vtrue$  but by the requirement that its value is
near the electroweak scale} (condition S1).

We present our stability results in Fig.~\ref{stab-only-plot} 
where we show the 64 curves
corresponding to all possible signs of $\alpha_i(\La)$ and the SM curve.
The black middle curve describes the SM limit, while the non-standard bounds
consist of 4 tight groups of 16 curves each: the two 
upper (green) curves  provide the limit for
$\al{p}(\La) = +1 $ while the lower (red) ones for $\al{p}(\La) =- 1$.
The graphs show that $ \al{tp} > 0 $ tends to 
destabilize the electroweak vacuum, and that the
limits obtained for fixed $ \al p(\La) $ and $ \al{tp} (\La) $
are almost independent of the signs of the other $ \alpha_i(\La) $ (their influence is
illustrated by the width of the curves).

In spite of the complicated nature of the analysis performed here, it is worth 
to trace the way  in which $\al{tp}$ could influence the lower limit on the
Higgs-boson mass. The key point is the fact that $\ocal_{t\phi}$ modify the 
relation (\ref{ht_mass}) between the top-quark mass and its Yukawa coupling.
Expressing the Yukawa coupling $f$ through $\mt$ one obtains the
following form of the fermionic contribution, $T$, to the effective 
potential (\ref{effpot}):
\beq
T=\left(\sqrt{2}\frac{\mt}{v_0}\right)^2\pb^2\left[1 +
\sqrt{2}\al{tp}\frac{v_0}{\mt}\frac{\pb^2-v_0^2}{\La^2}+
\ocal\left(\al{tp}\frac{v_0^2}{\La^2}\right)\right]\,.
\label{T}
\eeq
Since the instability (where the effective potential is bending down) takes 
place for $\pb\gg v_0$ therefore effectively we obtain substantial enhancement
or suppression (depending on the sign of $\al{tp}$) factor of top-quark 
contribution. Another mechanism of enhancing the contribution from $\ocal_{t\phi}$ 
is the very large numerical factor in front of $\al{tp}$ in the
evolution equation of $\al{p}$. Because of this a larger $\al{tp}$ drives
$\al{p}$ to larger values thereby again requiring a larger $\mh$. Both effects
combine leading to the dependence on $\al{tp}$ illustrated in 
Fig.\ref{stab-only-plot}.

It is worth noticing that $ \al p > 0 $ whenever
the effective operator $ {\cal O}_\phi $ is generated 
through the tree-level exchange of a heavy scalar isodoublet in the fundamental
high-scale theory. This
scenario allows for a Higgs boson mass below the SM stability
limit, and if this happens to be the case experimentally, the result
would not only indicate the presence
of new physics, but would also suggest the type of new
physics {\em and} provide an upper bound on its scale.

\subsection{Triviality bounds and combined limits}

\begin{figure}
\centering
\includegraphics[width=12cm]{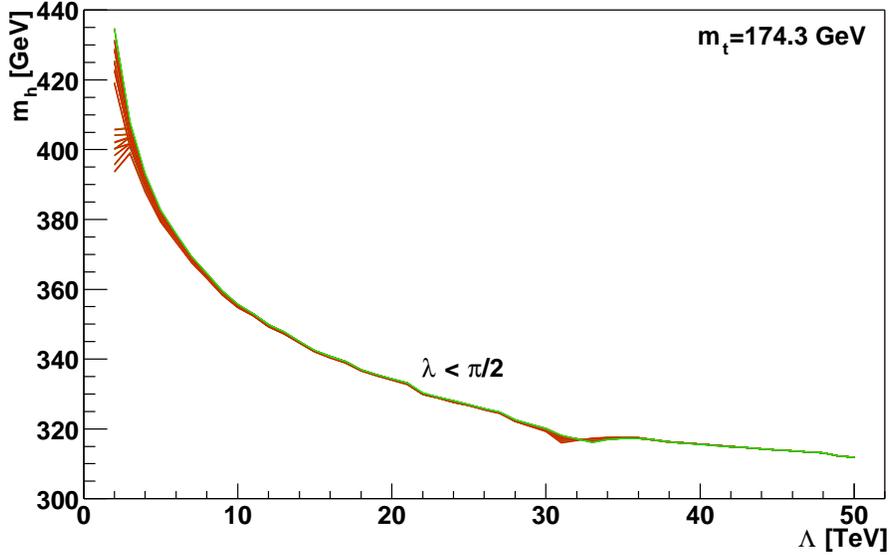}%
\caption{The upper bound on the Higgs boson mass from the standard triviality 
condition: $\la <\pi/2$ (the small-scale structure is due to numerical inaccuracies).}%
\label{triv-only-plot}%
\end{figure}

Turning now to the triviality bounds we first  note that the restrictions
imposed by T1 alone, presented in Fig. \ref{triv-only-plot} remain
unchanged in the presence of the effective operators. 
We include in this plot the SM result together with the 64 curves obtained
by taking $ \alpha_i(\La) = \pm1 $. 
It is seen that all 65 lines are nearly identical 
illustrating the fact that the SM upper limit stay approximately
unchanged in presence of the 
effective operators.

To qualitatively
understand this lack of sensitivity it is useful to consider the special
case where $ \alpha_{i \not= \phi} =0 $. In this case it follows form 
(\ref{beta_fun}) that $\al p(t)$ is a monotonically increasing function of
$t$~\footnote{ Here we consider heavy Higgs bosons, therefore $\la$ remains 
positive in the whole integration region,
it addition $f\gesim g,g'$ what guarantees that $d \log \alpha_\phi/ dt > 0$.};
and the numerical coefficients insure a rapid change from $ |\al p(\La)|=1 $
to $ |\al p ( \mz) | \lesim 0.1 $. Since $\al p $ below $ \La $ is small, its
presence does not significantly affect the evolution of $ \la $.
This is reinforced by the fact that the
$\alpha_i$-effects are always suppressed by small $\eta\equiv \la (v/\La)^2$
(a consequence of decoupling).
Therefore the corrections to the SM triviality bound from the non-standard physics 
(embedded in the coefficients $\alpha_i$) are negligible~\footnote{For strongly 
coupled new-physics corrections to this bound see~\cite{chan}.}.
Only for a very small scale $\La \lsim 3 \tev$ we observe slight deviations from the SM 
limit.
\begin{figure}
\centering
\includegraphics[width=\textwidth]{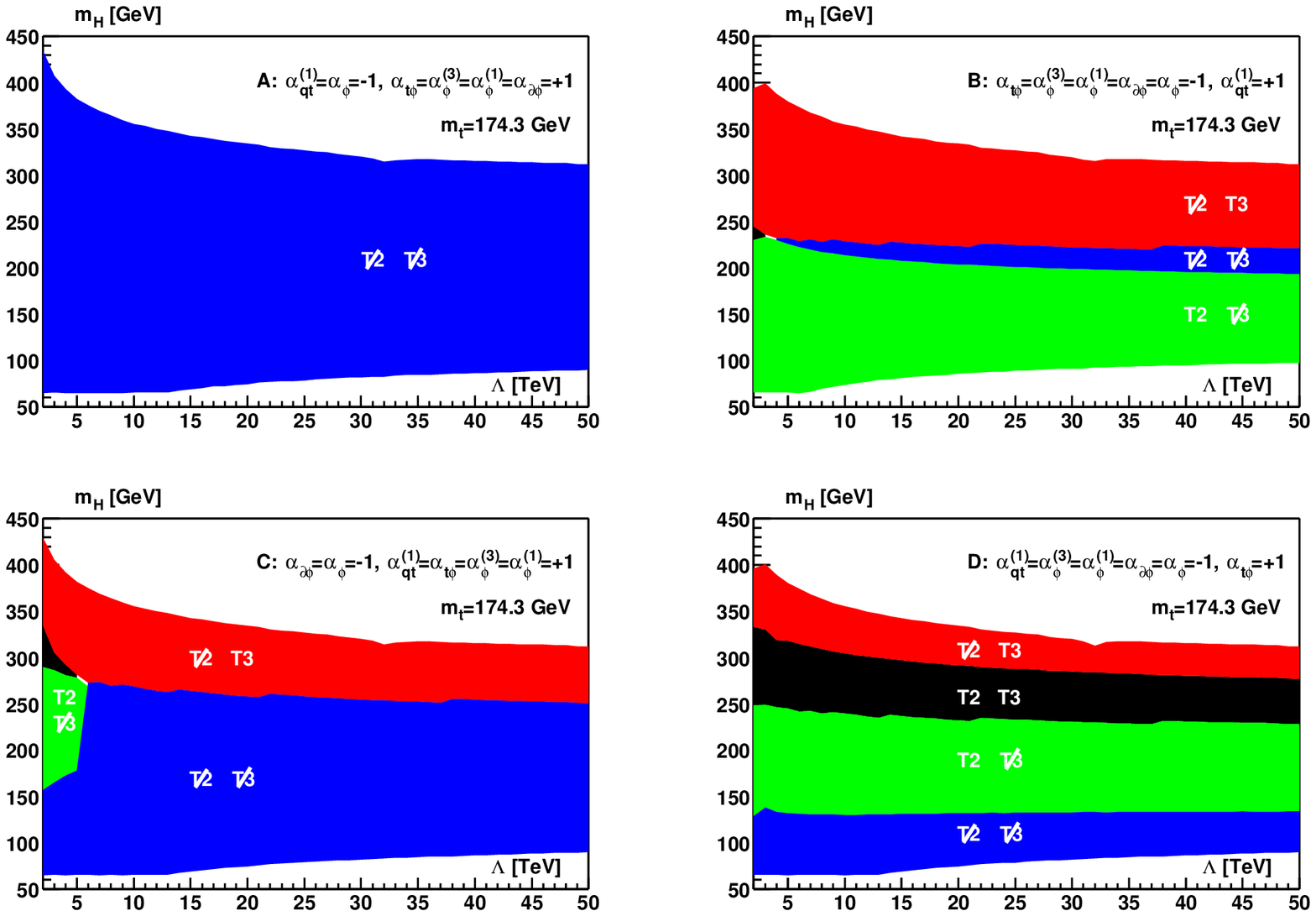}
\caption{The regions allowed and disallowed by the triviality conditions
T2 and T3 for 4 selected combinations of $\alpha_i$ specified in the main text
as models A, B, C, and D.
The black color regions, labeled T2 T3,
 are such that both conditions are satisfied;
red areas, labeled $\not\!\!\hbox{T2}$ T3, represent the regions where
T3 is satisfied but T2 is not;
green areas, labeled T2 $\not\!\!\hbox{T3}$, represent the regions where
T2 is satisfied but T3 is not; and blue areas,
labeled $\not\!\!\hbox{T2}$ $\not\!\!\hbox{T3}$, represent the regions where
neither T2 nor T3 are satisfied.}
\label{models-plot}
\end{figure}

The full triviality conditions for this model, however, also requires the imposition
of the conditions T2 and T3 and this leads to a rich structure and  manifold
possibilities. We have picked 4 illustrative cases~\footnote{As 
the triviality conditions 2 and 3 are symmetric under 
change of sign of all $\alpha_i$ we picked models with $\alpha_\phi=-1$.} 
with the following parameters: 
\beq
\begin{tabular}{|c|c|c|}
\hline
Case & $-1$ & $+1$  \cr\hline\hline
A & $\al{qt1},~\al{p}$ & $\al{tp},~\al{p3},~\al{p1},~\al{dp}$ \cr\hline 
B & $\al{tp},~\al{p3},~\al{p1},~\al{dp},~\al{p}$ & $\al{qt1}$ \cr\hline 
C & $\al{dp},~\al{p}$ & $\al{qt1},~\al{tp},~\al{p3},~\al{p1}$ \cr\hline 
D & $\al{qt1},~\al{p3},~\al{p1},~\al{dp},~\al{p}$ & $\al{tp}$ \cr\hline 
\end{tabular}
\eeq

We have restricted our analysis only to values of $ \La$
and $ \mh $ satisfying all the
stability conditions and the triviality condition T1 as 
presented on Figs.~\ref{stab-only-plot} and Fig.~\ref{triv-only-plot}
respectively.
The combined plots (containing both the lower and the upper bounds) 
corresponding to the models defined above  are 
presented in Fig.~\ref{models-plot} where we specify the additional
regions excluded by conditions T2 and T3.

The violation of condition T2 is always due to
$\al{p}$ or $\al{tp}$. For
large $\lambda$ (corresponding to large $m_h$)
T2 is similar but stronger than
condition T3b, while the opposite is true
for small $\lambda$ (i.e. small $m_h$). As a result
the $\not\!\!\hbox{T2}\hbox{T3}$ region (red) 
is always next to the upper edge of the region allowed by the triviality condition
($\la=\pi/2)$.

The fact that $\hbox{T2}\not\!\!\hbox{T3}$ region (green) 
or $\not\!\!\hbox{T2}\not\!\!\hbox{T3}$ region (blue)
 appears always adjoint to the lower edge (where $\la$ is
relatively small) is also easy to understand, for in this case
T3 is stronger, and therefore easier to violate.

Note that there exist models (e.g. the model A) such that 
either of the} requirements
T2 and T3 exclude the whole region between the SM-like upper limit ($\la < \pi/2$)
and the lower (stability) bound, which is due to large
contributions to the $\beta$-functions,
the tree-level potential and to the Higgs-boson mass from the effective operators. 
This illustrates the importance of conditions T2 and T3 that can
completely eliminate certain models, and severly limt the allowed
values of $ \La$ and $ \mh$ in others. These restrictions cannot be 
obtained using only the standard triviality conditions T1.

As we have already discussed, for the vacuum stability limits only $\al{p}$ and 
$\al{tp}$ were relevant. in contrast, \mem{the generalized (caused by T2 and/or T3) }
triviality limits depend on
{\em all} the $ \alpha_i $ since none of them plays a preferred}
role in the RG equations, which leads to
the observed rich texture. Note for instance the difference
between plots corresponding to models A and C for which $\al{p}$ and $\al{tp}$ 
are identical.

It is also worth emphasizing that condition T3 is equivalent to the
requirement that the effective operators generate small changes
in the $\beta$-functions, the tree-level potential and the Higgs mass. This
however, is not relevant for models where there is no relation
between the SM and the effective couplings (this would be similar
to the top contributions within the SM that need not be
small compared to those generated by the gauge fields or the
scalars). For such models the regions in Fig. \ref{models-plot}
labeled $\not\!\!\hbox{T3}$ are no longer forbidden.
It should also be mentioned that the actual strength of the triviality 
conditions T1, T2 and T3 is to certain extent arbitrary
({\em e.g} we could demand deviations below 20\%\ insted of the 25\%\
we used), and therefore 
shape of the allowed regions showed in Fig.\ref{models-plot} could 
be slightly modified if other conditions were specified.


\section{Summary and Conclusions}
\label{summary}

We have considered restrictions on the Higgs-boson mass that emerge
form requirement of perturbative behavior of coupling constants
(the triviality bound) and from the condition of stable electroweak 
vacuum,
taking into account possible non-standard interactions (of a typical scale $\La$) 
described by effective operators of dimension $\leq\;6$.   
The allowed regions in the ($\La,\mh$) space resulting from the stability and
triviality requirements has been determined and discussed in detail, taking
into account all the necessary constraints from precision 
tests of the Standard Model.
It was shown that for the scale of new physics in the region 
$2 \tev \lesim \La \lesim 50 \tev$
the Standard Model triviality upper bound (defined as an upper limit for 
the quartic coupling constant $\la$) remains unchanged,
whereas the lower bound from the requirement of vacuum stability could be  
substantially modified, depending on values of the coefficients of two dim~6 
operators: $\ocal_\phi=\inv3 | \phi|^6 $ and 
$\ocal_{t\phi}=| \phi |^2 \left( \bar q \tilde\phi t + \hbox{h.c.} \right)$. 
A natural generalization of the triviality
condition leads also to a substantial reduction of the allowed region in 
the ($\La,\mh$) space.

All the above considerations are applicable for the case where the
heavy physics is weakly coupled and decoupling, and has a
particle content that naturally generates the various operators considered,
especially $ \ocal_\phi$ and $ \ocal_{t\phi}$. However, there are models where these
operators are not generated~\footnote{$ \ocal_\phi$ appears 
at tree-level only in models containing heavy scalars of isospin $ \le3/2$;
$ \ocal_{t\phi} $ appears at tree-level only if the model contains 
a combination of heavy fermions and scalars of isospin $\le1$.} 
in which case the stability and triviality
bounds relax to those of the SM.

Finally we would like to mention that  several
discrepancies between
the results presented here and those of~\cite{Grzadkowski:2001vb}.
This is due to a series of typographical errors in~\cite{Grzadkowski:2001vb}: the signs
of the $ \alpha_i$ in the Lagrangian and effective potential (equations 
1,3 and 5 of that  reference) should be changed. Then the sign
convention for $ \alpha_i $
used in the present paper is {\em opposite} to the one used 
in~\cite{Grzadkowski:2001vb}; correspondingly the
sign of $ \alpha_\phi $ induced in a 2-higgs doublet model is
positive. In addition the triviality graph presented
in~\cite{Grzadkowski:2001vb} refers only to the usual SM condition, labeled T1 above;
the claims made in~\cite{Grzadkowski:2001vb} concerning the requirement T2
are incorrect.

\vspace*{0.6cm}
\centerline{\bf ACKNOWLEDGMENTS}

\vspace*{0.3cm}
This work is supported in part by the State
Committee for Scientific Research (Poland) under grant 5~P03B~121~20
and funds provided by the U.S. Department of Energy under grant No.
DE-FG03-94ER40837. BG and JP are indebted to
U.C. Riverside for the
warm hospitality extended to them while this 
work was being performed.

\vspace*{1cm}
\centerline{\bf APPENDIX}
\vspace*{1cm}

The issue of the vacuum stability in presence of non-standard physics
has been recently addressed by the authors of Ref.~\cite{Burgess:2002tj}. 
They conjectured that if the ground state of the
underlying theory has flat directions (no quartic interactions for
certain field configurations), then the effective theory will be
non-polynomial and an expansion in powers of light fields is not
justified. However, in the absence of fine tuning 
there will be
no flat directions, and these complications
do not arise. This is the situation considered
in the present paper where the
effective potential is by its construction
polynomial in light degrees of freedom.

As an illustration of their statement the authors of ~\cite{Burgess:2002tj}
provide an model containing two real scalar fields,
$\phi$ and $\Phi$, with the following potential
\beq V(\phi,\Phi)=-{1\over 2}m^2\phi^2+{1\over
8}\la\phi^4+{1\over 2}M^2\Phi^2 +\xi
\phi^3\Phi+\kappa\phi^2\Phi^2\ . 
\label{pot} 
\eeq
Assuming $M^2\gg m^2 > 0$, $V$ will have a minimum 
at $\phi^2 \simeq 2m^2/\la$ and $ \Phi  \simeq 0$
which will be stable provided
\beq 
\kappa>0\ ,\;\;\;\; {\rm
and}\;\;\;\; \la \geq 2 \frac{\xi^2}{\kappa}\,. 
\label{positivity}
\eeq

On the other hand in the effective theory obtained
by integrating out the heavy field $\Phi$ one obtains the following
low-energy potential:
\beq
 V(\phi)=-{1\over 2}m^2\phi^2+{1\over 8}\la\phi^4-{1\over
2}\xi^2{\phi^6\over M^2+2\kappa\phi^2}\,. 
\label{effpot-exa} 
\eeq
The authors of Ref.~\cite{Burgess:2002tj} then claim that expanding in
powers of $\phi^2/M^2$ leads to the result that
the potential (\ref{effpot-exa}) is unstable for parameters that at the same
time satisfy the positivity constraints (\ref{positivity}) 
for the underlying theory (\ref{pot}). Their conclusion
is then that stability requirements obtained using the
low-energy effective theory are inaccurate and can generate
much stronger bounds than those obtained using the full Lagrangian.

In order to investigate the effective Lagrangian derived from 
(\ref{effpot-exa}) we note that the large $M$
expansion is justified provided $\kappa\phi^2/M^2 \ll 1$. 
Adopting this restriction, one can easily find that
the effective theory is stable  when 
\beq
\la \geq 2\frac{\xi^2}{|\kappa|}\, 
\label{ours}.
\eeq
We thus reproduce the second stability 
condition in (\ref{positivity}); it 
is clear, however, that to the lowest order 
in $\kappa\phi^2/M^2 $ we cannot 
recover the constraint $ \kappa > 0 $. This is so because
(\ref{ours}) is the result of physics at or below the 
cutoff $ \Lambda = M/\sqrt{|\kappa|}$.
In contrast, the condition on $ \kappa$ results from 
of some physics (or fine tuning) above the cutoff, and cannot be obtained using
the effective theory. 
It should also be emphasized here that the
effective Lagrangian approach presupposes that 
all terms allowed by the symmetries of the model are
present in the original Lagrangian, which is not the
case for (\ref{pot}); accordingly, the constraint
on $ \kappa $ is significantly modified if we
allow a term $ \propto \Phi^4 $. Though one cannot
draw any general conclusions experimenting with a fine tuned
potential such as (\ref{pot}), this example does
provide a useful illustration of the implications of the 
naturality assumption in effective theories.

\vskip 1cm

\end{document}